\pdfoutput=1

\documentclass[11pt]{article}

\usepackage{naacl2021}

\usepackage{times}
\usepackage{latexsym}

\usepackage[T1]{fontenc}

\usepackage[utf8]{inputenc}

\usepackage{microtype}

\usepackage{tikz}
\usepackage{array}
\usepackage{calc}
\usetikzlibrary{positioning}
\usetikzlibrary{calc}
\usepackage{float}
\usepackage{amsmath} 
\usepackage{listings}
\usepackage{xcolor}

\usepackage{pifont}

\definecolor{stringcolor}{RGB}{255,105,180}  
\definecolor{commentcolor}{RGB}{87,166,74}   
\definecolor{keywordcolor}{RGB}{86,156,214}  
\definecolor{backgroundcolor}{RGB}{30,30,30} 
\definecolor{textcolor}{RGB}{212,212,212}    

\usepackage{booktabs}
\usepackage{multirow}

\title{Cerebrum (AIOS SDK): A Platform for Agent Development, Deployment, Distribution, and Discovery}

\author{Balaji Rama \\
  \texttt{balaji.rama@rutgers.edu}\And
  Kai Mei\\
  \texttt{kai.mei@rutgers.edu}\AND
  Yongfeng Zhang \\
  \texttt{yongfeng.zhang@rutgers.edu}
}

\begin{document}
\maketitle
\begin{abstract}
Autonomous LLM-based agents have emerged as a powerful paradigm for complex task execution, yet the field lacks standardized tools for development, deployment, distribution and discovery of agents. We present Cerebrum, an Agent SDK for AIOS that addresses this gap through three key components: (1) a comprehensive SDK featuring a modular four-layer architecture for agent development, encompassing LLM, memory, storage, and tool management; (2) a community-driven Agent Hub for sharing and discovering agents, complete with version control and dependency management; (3) an interactive web interface for testing and evaluating agents. The platform's effectiveness is demonstrated through implementations of various agent architectures, including Chain of Thought (CoT), ReAct, and tool-use agents. Cerebrum advances the field by providing a unified framework that standardizes agent development while maintaining flexibility for researchers and developers to innovate and distribute their agents. The live website is at \href{https://app.aios.foundation}{https://app.aios.foundation}, the code is at \href{https://github.com/agiresearch/Cerebrum}{https://github.com/agiresearch/Cerebrum}, and video \href{https://app.aios.foundation/video-demo}{https://app.aios.foundation/video-demo}.
\end{abstract}

\section{Introduction}

Autonomous LLM-based agents (agents for short) have emerged as a transformative paradigm in applying and advancing the capabilities of Large Language Models (LLMs) beyond text prediction to executing complex tasks through planning, reasoning, tool using, and goal-directed actions \cite{ge2024openagi,shinn2024reflexion,li2023camel,deng2024mind2web,mei2024aios}. The paradigm scales to real-world issues such as web browsing \cite{iong2024openwebagent,deng2024mind2web}, social simulation \cite{park2023generative,pang2024self}, and decision-making \cite{hua2024game,mao2023alympics}.


Although with the fast advancement of LLM-based agent research in the recent year, there still lacks a unified platform for researchers and developers to develop, deploy, and distribute their agents, and for users to discover and use the agents. This demonstration paper introduces \textit{Cerebrum} AgentHub, which not only provides an SDK for agent development and deployment, but also provides a web-based platform for agent developers to share their agents, and for agent users to easily use the agents both through interactive web-based UI interface and through code-based calling of pre-loaded agents through one line of code.

More specifically, \textit{Cerebrum} is a library accompanied with a live demo dedicated to supporting a standardized way to build, run, deploy, and distribute agents and agent components. At the core of the library is a unified framework for constructing diverse agents, containing optimized implementations of popular agent methodologies such as Chain of Thought (CoT) \cite{wei2022chain} and ReAct \cite{yao2022react}, with the goal of supporting implementations of agent variants that are easy to read, extend, and deploy. Furthermore, the library supports the distribution and usage of user-created agents in a centralized agent hub.

\section{Related Work}
AI Agents have long been considered an important step towards generalist intelligence \cite{wooldridge1995intelligent,jennings1998roadmap,bresciani2004tropos}. Acting as core coordinators, agents are envisioned as intelligent entities that can perceive their surroundings, build memories \citep{xu2025mem, wang2024agent}, devise plans, and autonomously carry out actions to fulfill tasks \cite{wang2024survey,deng2024mind2web, shi2025from, xu2025instructagent, zhang2024agent}. 

The emergence of LLMs has drastically expanded the potential for advancing agent technology, as demonstrated by recent breakthroughs \cite{liu2024agentlite,zhang2024agentohana}. Traditional prompt-based interactions are typically static, functioning as direct input-output exchanges with limited adaptability. In contrast, LLM-driven agents are designed to enable dynamic decision-making processes, granting them the ability to interpret context, generate flexible responses, and act independently \cite{shinn2024reflexion}. This evolution allows agents to transition from handling simple, single-step tasks to becoming versatile, general-purpose problem solvers \cite{ge2024openagi}.

\section{Cerebrum Library Design}
Cerebrum is designed to provide a standardized framework for developing LLM-based agents, addressing the growing need for systematic agent architectures in the artificial intelligence community. The library implements a modular approach that facilitates both research and production deployments while maintaining flexibility for various use cases. 
The library's architecture consists of two primary components: (1) a layered system for agent composition and (2) a client interface for kernel communication. This dual architecture enables both fine-grained control over agent behavior and rapid development through high-level abstractions.

\subsection{Layer Architecture}
\begin{figure*}[h]
    \centering
    \includegraphics[width=\textwidth]{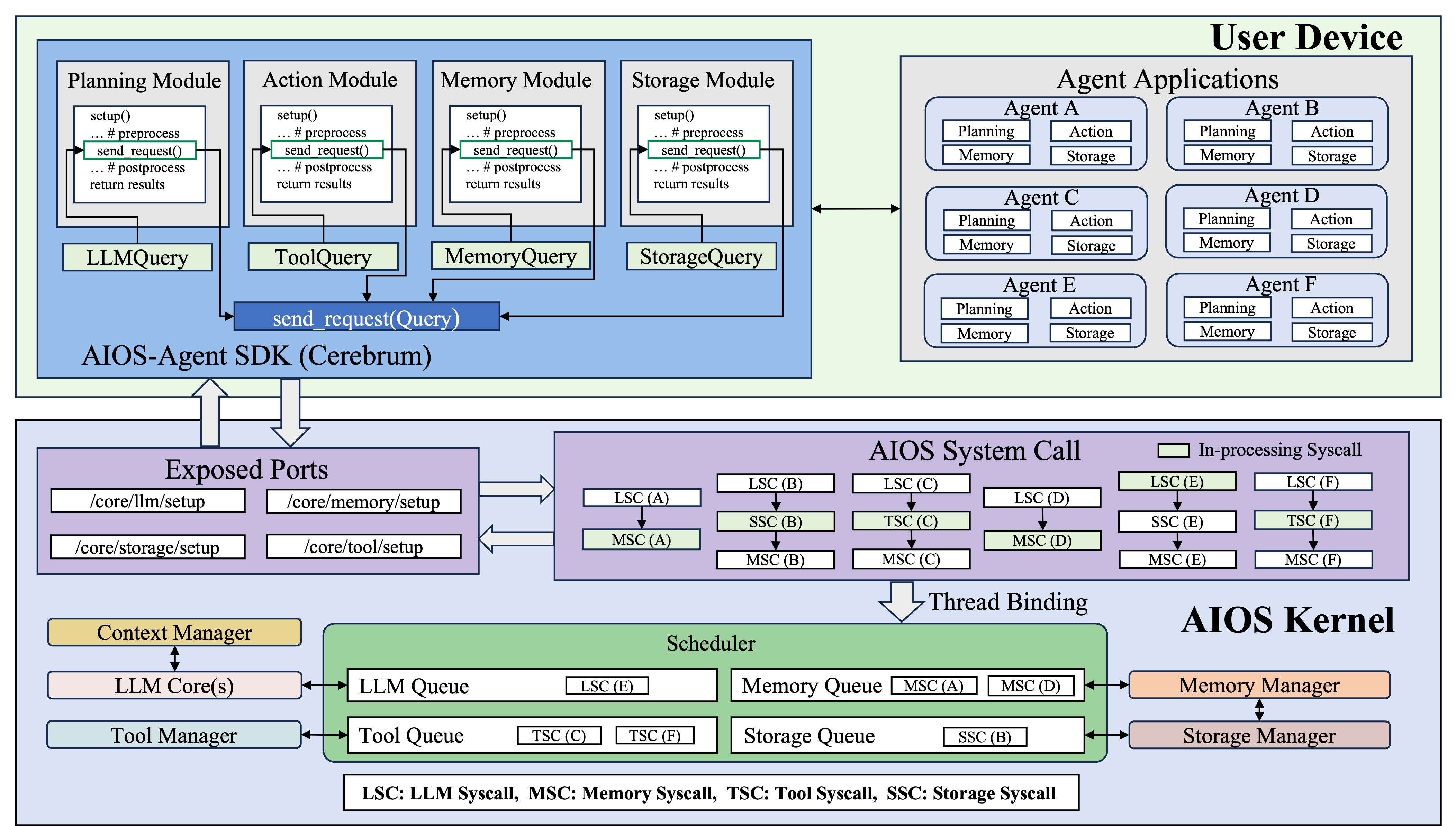}
    \caption{The architecture of Cerebrum on the basis of AIOS. }
    \label{fig:architecture}
    \vspace{-15pt}
\end{figure*}

Every agent in the library is fully defined by four foundational building blocks shown in the diagram in Figure~\ref{fig:architecture}: (a) an LLM Layer, which manages model interactions and resource allocation, (b) a Memory Layer, which handles context management and state persistence, (c) a Storage Layer, which provides durable storage capabilities, and (d) a Tool Layer, which enables structured interaction with external systems. Most agent development requirements can be addressed through the composition of these four components. These layers represent connections with the corresponding parts of the AIOS kernel \citep{mei2024aios, mei2025eccos, ge2023llm}, allowing agents to run on the kernel.

\subsubsection{Large Language Model Layer}
The Large Language Model (LLM) layer allows agents to utilize LLM cores as their backbones. While each supported model may have unique characteristics, the LLM Layer provides a standardized interface that enables seamless switching between different providers and architectures. Cerebrum, for the most part, is able to determine smart defaults for LLM parameters such as temperature, resource constraints, etc, but also provides additional fine-grained control over these parameters. 

\subsubsection{Memory Layer}
The Memory Layer implements sophisticated working memory management for agents, crucial for maintaining context and enabling informed decision-making. The layer provides configurable memory limits, eviction strategies, and custom policy support. Memory limits are specified in bytes, with default configurations suitable for most applications. It implements an LRU-k eviction approach, where k determines the number of items considered for removal when memory limits are reached. This enables agents to maintain relevant context while efficiently managing computational resources through configurable eviction policies. 

\subsubsection{Storage Layer}
The Storage Layer provides persistent storage capabilities essential for long-term knowledge retention and cross-session continuity. The system supports both traditional hierarchical storage through a root directory structure and modern vector databases for efficient similarity-based retrieval. When vector databases are enabled, the system can be configured with specific embedding models, dimension parameters, and indexing strategies. This flexibility allows for optimization based on specific deployment requirements and enables sophisticated knowledge management capabilities.

\subsubsection{Tool Layer}

The tool layer implements a comprehensive interface that handles the complexities of tool discovery, loading, and integration with large language models. Through a standardized protocol, it manages tool initialization, parameter validation, and execution flow while maintaining proper error handling.

\subsubsection{Overrides Layer}
Cerebrum features an optional Overrides Layer that provides fine-grained control over AIOS Kernel parameters. While most standard deployments operate effectively with default configurations, this layer enables advanced customization (e.g., scheduler) for specialized use cases. Modifications can only be performed through carefully designed interfaces. In this way, the modifications made by the developers will not influence the system stability.

\begin{figure*}[h]
    \centering
    \includegraphics[width=\textwidth]{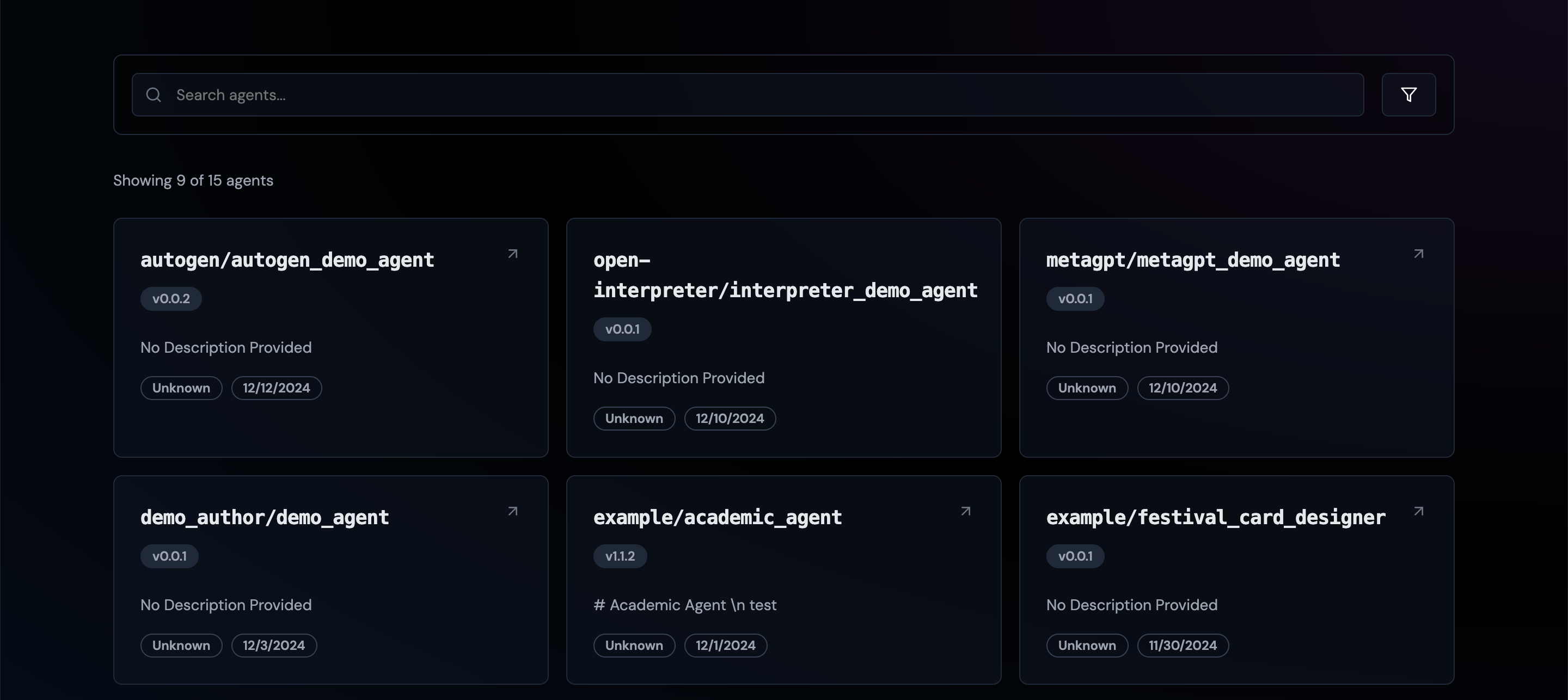}
    \caption{Cerebrum AgentHub Demo: \url{https://app.aios.foundation/agenthub}}
    \label{fig:your-label}
    \vspace{-15pt}
\end{figure*}

\subsection{Manager Module}



A key component of the library is the \textit{manager} abstraction, which orchestrates agent and tool lifecycle operations. The manager system consists of two specialized components: the agent manager and the tool manager. These managers share largely identical functionality, with minor differences in their handling of their respective artifacts. Their primary responsibilities encompass distribution, versioning, caching, packaging, downloading and uploading.

\subsubsection{Distribution, Version Control, Dependency Resolution}
The framework maintains a centralized repository for agents, supporting versioning and dependency management. Each agent is uniquely identified by a triple of {author, name, version}, enabling precise version control and reproducibility of agents.

\subsubsection{Caching, Packaging}
To optimize system performance and minimize network overhead, the Cerebrum manager designs and implements the caching mechanism for both tools and agents. The packaging provides bundling of agents and their dependencies, allowing for reproducible deployments across different environments. 
And the cerebrum employs version-aware storage and retrieval strategies, enabling efficient management of different component versions while ensuring consistency across deployments. 

\subsubsection{Upload, Load, Download}
Cerebrum supports direct uploading of packaged .agent and .tool files to their respective hubs through a streamlined interface. The download functionality is provided with built-in verification and integrity checks. The loading of agents is dynamic This loading is dynamic, allowing agents to be instantiated and used at run-time while maintaining proper isolation, which ensures operational stability and prevents unintended interactions between different agents and tools. 

\subsection{Client Interface and Auto-configuration}
The library also features a client interface to both allow users and agents themselves to interact with the AIOS kernel, as well as the AIOS kernel to interact with agents. Users can use the client to run agents on the AIOS kernel, while AIOS uses the kernel to download, load, and run agents and tools. Additionally, we feature Auto- classes that abstract around multiple components of the Cerebrum library to allow for 1-2 line loading and deployment.

The client interface serves as the bridge between the application logic (layers + manager) and the AIOS kernel. The interface adopts a declarative configuration approach, where 
The client system implements a builder pattern that maintains strict initialization order dependencies while providing a fluent interface for component composition.

The client interface is augmented by a set of Auto classes that provide factory methods for reusable agent components. Similar to the Auto classes in the Transformers library, these components handle the complexities of initialization with sensible defaults and full configurability.

\begin{lstlisting}
# Load the agent
agent = AutoAgent.from_preloaded("example/academic_agent")

# Use the agent
response = agent.run({
    'task': "Your input here"
})
\end{lstlisting}

\section{Agent Standard}
The library provides a comprehensive agent building framework that emphasizes composition over inheritance. Unlike traditional agent frameworks that often require deep understanding of implementation details and rigid structual foundations, Cerebrum's building system enables rapid development through high-level abstractions while maintaining access to low-level controls when needed.

The framework introduces the concept of agent specifications - declarative definitions that describe an agent's capabilities, resource requirements, and behavioral patterns. These specifications can be composed, extended, and modified dynamically, enabling flexible agent architectures that can adapt to different deployment scenarios. Its implemented resource management strategies include:
\ding{172} Automatic tool resolution and dependency management. 
\ding{173} Dynamic resource allocation based on agent specifications.

A key innovation in the building framework is its handler system, which provides extension points for customizing agent behavior without modifying core components. These handlers can intercept and modify agent operations at various stages, enabling behavior patterns while maintaining the benefits of the standard agent lifecycle.

\begin{figure*}[h]
    \centering
    \includegraphics[width=\textwidth]{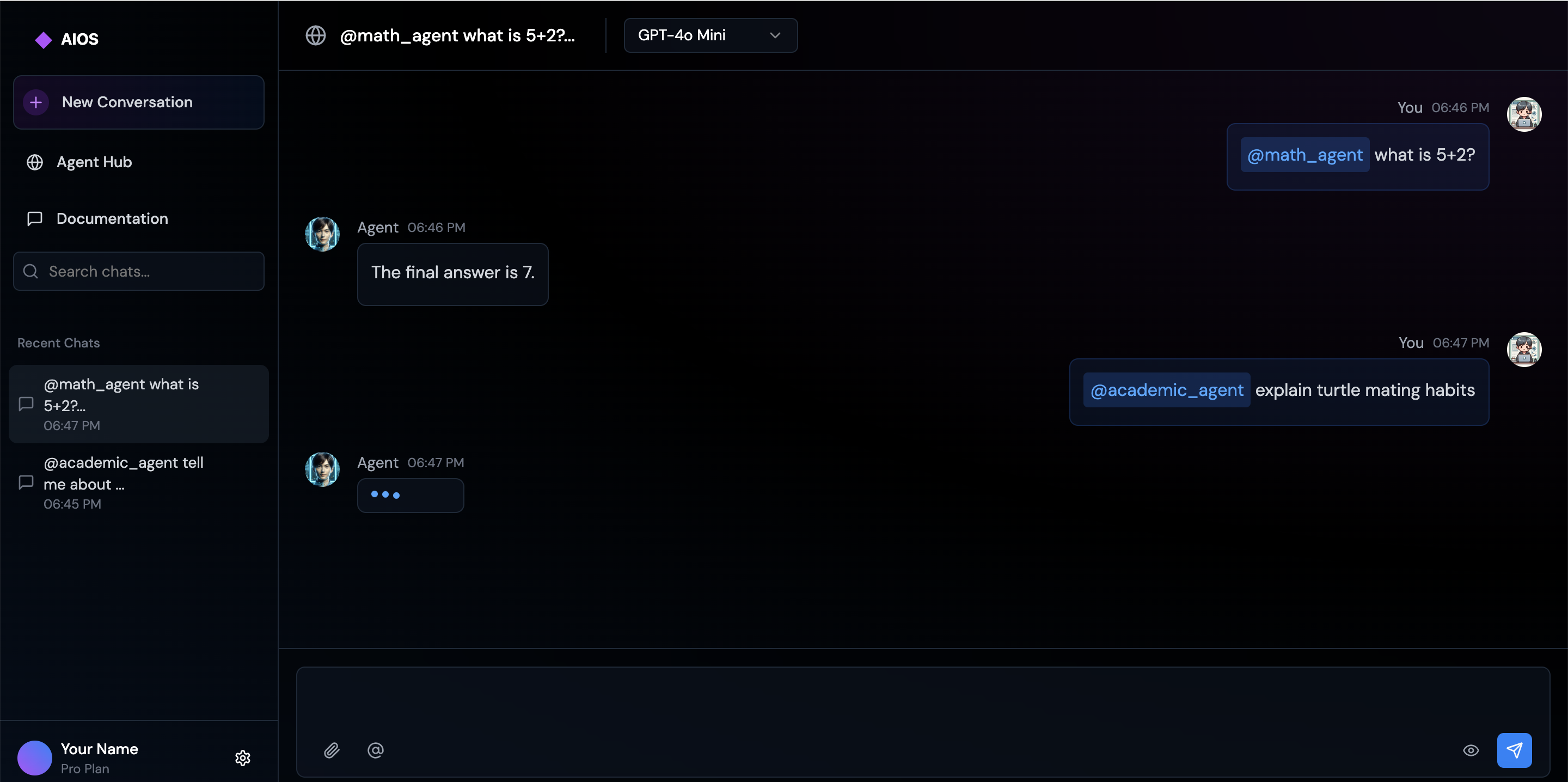}
    \caption{AgentChat Page: \url{https://app.aios.foundation/chat}}
    \label{fig:agent_detail}
    \vspace{-15pt}
\end{figure*}

\section{Community Agent Hub}
\textit{Cerebrum} implements an open-source distribution platform for AI agents, following a model similar to Hugging Face's hub architecture. The Community Agent Hub serves as a central repository where researchers can freely share, discover, and utilize agents that conform to the Cerebrum agent specification. The hub itself is a hosted, publicly accessible server featuring both an overall listing of all agents and agent-specific pages. 

Agents and tools are stored in an encrypted, hashed, and compressed format, containing references to their individual component files. Individual agent landing pages provide comprehensive information including:
\ding{172} Version control and release history. 
\ding{173} Direct access to the agent's inference API endpoints via Agent Chat. 
\ding{174} Licensing information and README documentation. 
\ding{175} Source code accessibility for transparency and reproducibility. 
\ding{176} Usage instructions.

A current limitation of the hub is the absence of a formal vetting process for uploaded agents. 
Future work may explore implementing security scanning, performance validation, and compliance checking mechanisms.

\begin{figure*}[h]
    \centering
    \includegraphics[width=\textwidth]{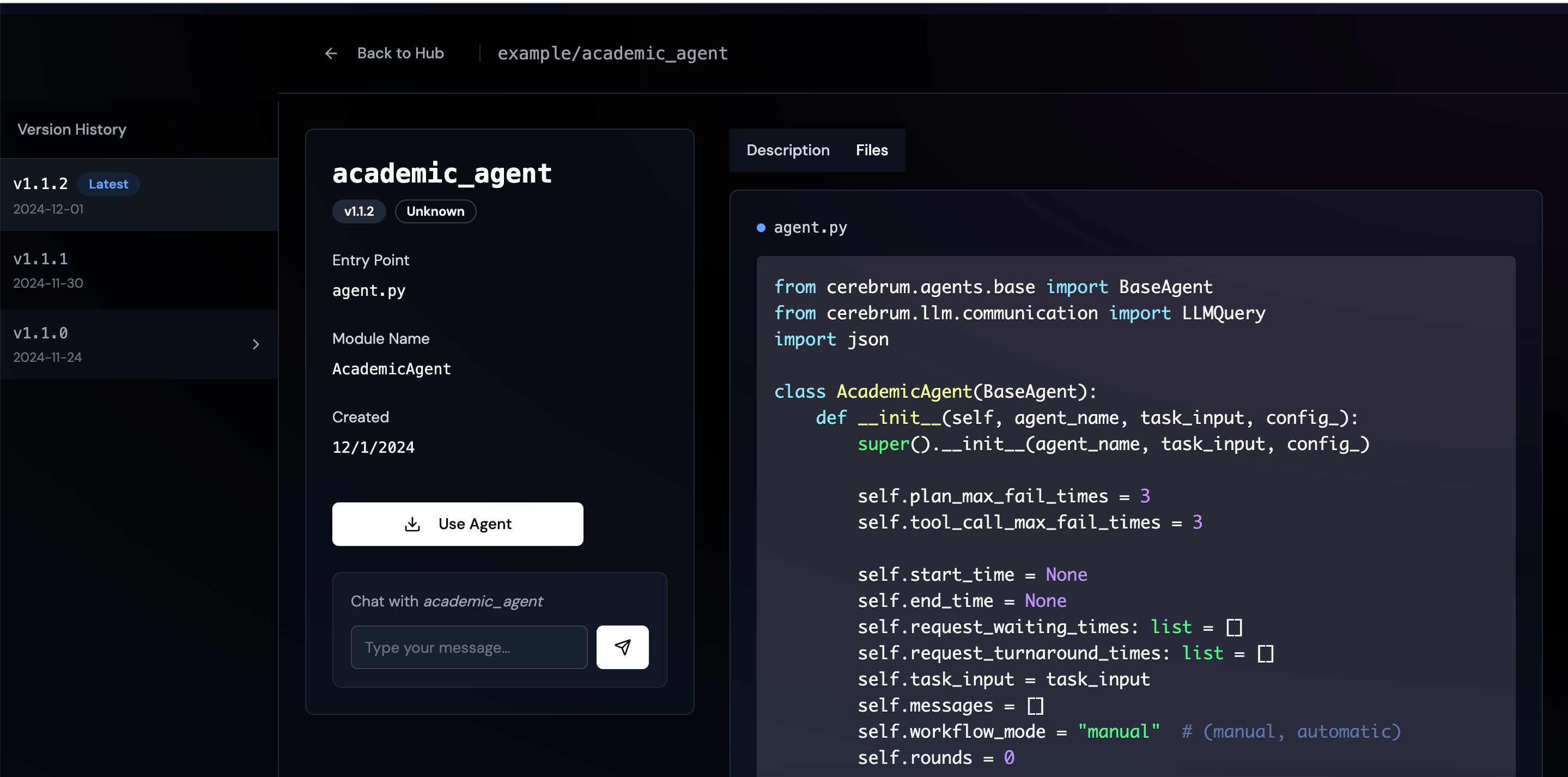}
    \caption{Agent Details: \url{https://app.aios.foundation/agents/example/academic_agent}}
    \label{fig:agent_detail}
\end{figure*}

\section{Community Agent Chat}
To facilitate direct interaction with and evaluation of agents, we provide a public chat interface that enables real-time communication with agents hosted in the Community Hub. This interface serves as both a research tool for analyzing agent behavior and a demonstration platform for agent capabilities.

The chat system implements a mention-based interaction model using the @ syntax (e.g. name query), where users can invoke specific agents. While the current implementation supports single agent interactions, multiple agents can be interacted with in a single conversation. Rate-limiting mechanisms are in place to ensure system stability and fair resource allocation.

Users can interact with any agents stored in the agent hub, which are served on a remote AIOS instance. The agent chat functions as a wrapper around a Cerebrum client that communicates with this remote instance. The system supports persistent, on-device chat and conversation memories, allowing users to maintain multiple different chats, which can also be deleted as needed.

\section{Applications}
To demonstrate the real-world applications of Cerebrum, we implemented four distinct agents that showcase different prompting techniques and capabilities: Chain of Thought (CoT), ReAct, a baseline chatbot, and a tool-augmented agent. These implementations serve to validate the flexibility and expressiveness of our agent specification framework.

\subsection{Baseline Chatbot}
To establish a performance baseline, we implemented a standard chatbot agent that maps input directly to output without intermediate reasoning.

\begin{equation}
P(y|x) = \text{LLM}(\text{prompt}(x))
\end{equation}

This serves as a control for evaluating the benefits of more sophisticated prompting techniques.

\subsection{Chain of Thought Agent}
Chain of Thought prompting \cite{wei2022chain, wang2022self, jin2024impact} enables step-by-step reasoning in language models. The process can be formalized as follows:

Given input query $x$, the agent generates intermediate reasoning steps $s_1, \ldots, s_n$ before producing final output $y$:
\begin{equation}
\small
P(y|x) = \sum_{s_1,\ldots,s_n} P(y|s_n)P(s_n|s_{n-1})\ldots P(s_1|x)
\end{equation}

The prompt template is implemented as:
\begin{equation}
\small
\text{prompt}(x) = \text{``Let's approach this step by step:''} + x
\end{equation}

Each reasoning step $s_i$ is explicitly generated and tracked, allowing for:
\ding{172} Verification of logical consistency. 
\ding{173} Identification of failure points.
\ding{174} Analysis of reasoning patterns. 

\subsection{ReAct Agent}
ReAct \cite{yao2022react} combines reasoning and action in an interleaved manner. We implement this as a Markov Decision Process where:

\begin{itemize}
    \item State space $S$: Current context + reasoning history
    \item Action space $A$: \{Thought, Action, Observation\}
    \item Transition function $T(s'|s,a)$: Updates state based on chosen action
    \item Policy $\pi(a|s)$: Determines next action given current state
\end{itemize}

The agent follows the cycle:
\begin{equation}
\scriptsize
\text{Thought} \xrightarrow{\text{leads to}} \text{Action} \xrightarrow{\text{generates}} \\
\text{Observation} \xrightarrow{\text{informs}} \text{Thought}
\end{equation}

Formally, at each step $t$:
\begin{align}
    a_t &\sim \pi(\cdot|s_t) \\
    s_{t+1} &= T(s_t, a_t)
\end{align}

\subsection{Tool-Augmented Agent}
The tool-augmented agent demonstrates Cerebrum's external tool integration capabilities. The agent employs a hierarchical decision process:

\begin{enumerate}
    \item Tool Selection: $$P(\text{tool}|x) = \text{softmax}(f_{\text{select}}(x))$$
    \item Tool Parameter Generation: $$\text{params} = f_{\text{params}}(x, \text{tool})$$
    \item Tool Execution: $$\text{result} = \text{execute}(\text{tool}, \text{params})$$
    \item Response Generation: $$y = f_{\text{respond}}(x, \text{result})$$
\end{enumerate}

Where $f_{\text{select}}$, $f_{\text{params}}$, and $f_{\text{respond}}$ are learned functions implemented via prompt engineering.

\section{Conclusion}
\vspace{-5pt}
We presented Cerebrum, a platform for developing, deploying, and distributing LLM-based agents. The platform addresses fundamental challenges in the agent development ecosystem through three key innovations: (1) a modular four-layer architecture that standardizes agent development while maintaining flexibility, (2) a community-driven Agent Hub that facilitates agent sharing and discovery, and (3) an interactive chat interface for direct agent evaluation and testing. Our implementations of various agent architectures, including CoT, ReAct, and tool-augmented agents, demonstrate the platform's versatility and effectiveness.

Looking forward, we envision several directions for future work. First, enhancing the Agent Hub with formal security and performance validation mechanisms would increase trust and reliability in shared agents. Second, expanding the tool layer to support more complex multi-agent interactions and collaborative scenarios could enable more sophisticated agent behaviors. Finally, developing standardized benchmarks and evaluation frameworks specifically for testing agents built with Cerebrum would help quantify and improve agent performance across different architectures and use cases.

Through its open-source nature and emphasis on standardization, Cerebrum aims to accelerate research and development in the rapidly evolving field of LLM-based agents, while fostering a collaborative ecosystem for sharing and building upon existing agent implementations.

\bibliography{anthology,custom}
\bibliographystyle{acl_natbib}




\end{document}